\documentclass[a4paper]{article}
\usepackage[utf8]{inputenc}
\usepackage{Odyssey2022}
\usepackage{epsfig,amssymb,amsmath}
\usepackage{multirow}
\usepackage{newunicodechar}

\DeclareUnicodeCharacter{00A0}{ }

\ninept

\usepackage[
backend=biber,
style=ieee,
maxbibnames=3,
maxcitenames=3,
doi=false,isbn=false,url=false,eprint=false
]{biblatex} 

\defbibheading{bibliography}[\refname]{}

\renewcommand{\vec}{\boldsymbol} 

\title{Spoofing-Aware Speaker Verification with Unsupervised Domain Adaptation}

\makeatletter
\def\name#1{\gdef\@name{#1\\}}
\makeatother

\name{\em Xuechen Liu{$^1{}^,{}^2$}, Md Sahidullah{$^2$}, Tomi Kinnunen{$^1$}}
\address{
  {$^1$}School of Computing, University of Eastern Finland, Joensuu, Finland\\
  {$^2$}Universit\'{e} de Lorraine, CNRS, Inria, LORIA, F-54000, Nancy, France \\
  {\small \tt xuechen.liu@inria.fr}}
\begin{document}
\maketitle

\begin{abstract}
In this paper, we initiate the concern of enhancing the spoofing robustness of the automatic speaker verification (ASV) system, without the primary presence of a separate countermeasure module. We start from the standard ASV framework of the ASVspoof 2019 baseline and approach the problem from the back-end classifier based on probabilistic linear discriminant analysis. We employ three unsupervised domain adaptation techniques to optimize the back-end using the audio data in the training partition of the ASVspoof 2019 dataset. We demonstrate notable improvements on both logical and physical access scenarios, especially on the latter where the system is attacked by replayed audios, with a maximum of 36.1\% and 5.3\% relative improvement on bonafide and spoofed cases, respectively. We perform additional studies such as per-attack breakdown analysis, data composition, and integration with a countermeasure system at score-level with Gaussian back-end.
\end{abstract}
\section{Introduction}

\emph{Automatic speaker verification} (ASV) \cite{asv_2015} is one of the most natural and convenient ways for biometric person authentication. Moving on from conventional models such as \emph{Gaussian mixture models} (GMM) and \emph{i-vector}, state-of-the-art ASV systems based on deep neural networks (DNNs) show reasonably good recognition accuracy on different speech corpora such as NIST SREs, VoxCelebs, and SITW~\cite{asv-dnn-review_2021}. However, ASV performance is substantially degraded in the presence of spoofing attacks made with voice conversion, text-to-speech synthesis, and replay~\cite{wu2015spoofing,voicepad2019}. High vulnerability of ASV system has been demonstrated using the speech corpora developed for automatic speaker verification spoofing and countermeasures challenge (ASVspoof)~\cite{wu15e_interspeech,kinnunen17_interspeech,todisco19_interspeech,yamagishi21_asvspoof}.

To protect the ASV systems from spoofing attacks, spoofing countermeasures (CM) are required which can discriminate the natural or human speech from spoofed or computer-generated/playback speech~\cite{wu2015spoofing,voicepad2019}. By leveraging data resources such as ASVspoof challenges, the main focus of the community has been in developing a dedicated CM suitable for detecting a wide variety of spoofing attacks. Then, the separately designed CM is integrated with the ASV system at score-level or decision-level~\cite{sahidullah16_interspeech}. The work in~\cite{todisco19_interspeech} applied Gaussian back-end based fusion of ASV and CM scores. More recently, joint optimization of ASV and CM was performed on embedding space to detect both human and spoofed imposters~\cite{TIFS2021}. The combined system has shown substantial performance improvement when dealing with imposters using spoofed audio.

While the integrated systems have demonstrated better performance than standalone ASV, such system combination raises several concerns. First, this strategy requires additional workloads for design, training, and optimization. Second, the low generalizability of the CM system can have a severe impact on integrated systems even though ASV systems show relatively better generalization. Moreover, the presently used integration methods have their limitations and they often degrade overall ASV performance by increasing either false acceptance rate or false rejection rate for trials with bonafide speech. Recently the \emph{spoofing-aware speaker verification (SASV) challenge} has been announced \cite{sasv2022}, aiming at designing spoofing-aware ASV systems. However, the problem that emerged with CM persists.

Inspired by the above concerns, in this work, we initiate the need of making the ASV system itself more aware of spoofing attacks, without the primary presence of CM. We particularly focus on \emph{unsupervised domain adaptation} (DA) by considering the spoofed audio samples available for training CM systems. We hypothesize that the use of spoofed audio data during domain adaptation would provide better generalization for ASV systems to spoofed imposters. To the best of our knowledge, this is the first comparative study on unsupervised DA techniques aiming at improving the generalization power of the ASV system towards the needs of anti-spoofing.

\begin{figure*}[!ht]
  \begin{center}
    \centering
     {\includegraphics[width=14cm]{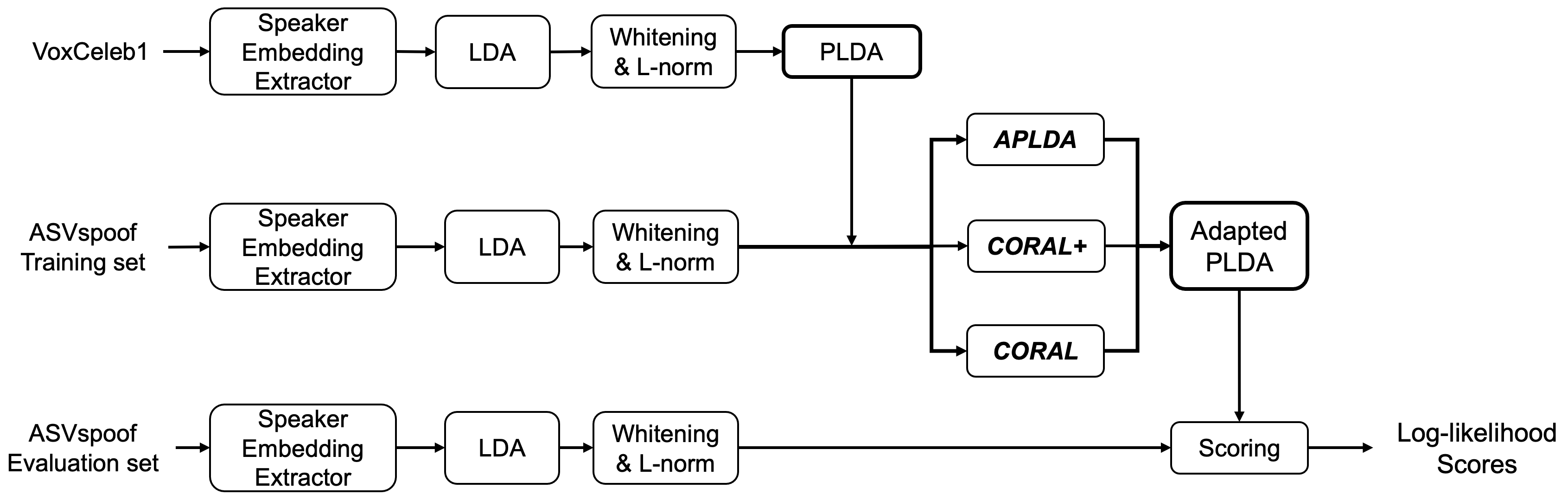}}
    \caption{An illustration of the back-end system, along with introduced unsupervised domain adaptation techniques.}
    \label{fig:methods}
  \end{center}
\end{figure*}

\section{Unsupervised Domain Adaptation}

One of the key challenges in speaker recognition is generalization across different domains. Whether due to channel, speaker population, language (or other) factors, mismatch in training and test data induces substantial performance drop. While this \emph{domain mismatch} problem can be alleviated in different ways, it is particularly convenient to update the back-end classifier to be better matched with the new domain. In this section, we thus describe the methods used to address the problem of creating spoofing-aware ASV, tackling the back-end system based on \emph{probabilistic linear discriminant analysis} (PLDA) \cite{plda_2006}, a Gaussian-based classifier that has been widely used for ASV.  

\subsection{PLDA}
PLDA models both channel and speaker variability through specifically structured covariance matrices. By denoting speaker embedding as $\vec{\phi}$, PLDA models the speaker embedding space as follows~\cite{plda_2006, coralplus, fastplda}:

\begin{align}
    p(\vec{\phi}) = \mathcal{N}(\vec{\phi} | \vec{\mu}, \mathbf{\Phi}_{b} + \mathbf{\Phi}_{w}),
\end{align}

where $\vec{\mu}$ is the global mean vector, $\mathbf{\Phi}_{b}$ and $\mathbf{\Phi}_{w}$ respectively model the between-class and within-class covariances. 

Since we use simplified PLDA~\cite{kaldi_aplda}, by denoting $\mathbf{F}$ and $\mathbf{G}$ as the speaker and channel loading matrices respectively, the two covariance matrices are structured as $\mathbf{\Phi}_{b} = \mathbf{F}\mathbf{F}^{T}$, $\mathbf{\Phi}_{w} = \mathbf{G}\mathbf{G}^{T} + \mathbf{\Sigma}$,
where $\mathbf{\Sigma}$ is a diagonal matrix which models residual covariances. For more information about the training and scoring procedure of PLDA, the readers are refer to~\cite{coralplus, plda, fastplda}.

\subsection{Domain Adaptation with ASVspoof}

The problem of domain mismatch remains for PLDA despite its promising performance in general, which needs the presence of in-place training data. Nevertheless, for complicated scenarios it might be hard to obtain enough amount of data to train a PLDA model.
Since the quantity of data from the new domain might be limited, a common practice on this issue is to first train a PLDA using out-of-domain (OOD) data and adapt it with a small quantity of in-domain (inD) data from the new domain. Depending on speaker labeling of the in-domain data, both supervised (speaker labels required) \cite{TIFS2021} and unsupervised (speaker labels not required) \cite{kaldi_aplda} methods are available. In this study, we focus on the latter not only because of its broader scope but also from a more fundamental perspective as our scenario involves the use of \emph{spoofed} audio-data in adaptation where speaker labels are not necessarily unambiguously defined. For instance, a VC-transformed speaker may `blend' voice characteristics of both source and target speakers. 

For the adaptation, we implement various methods with the bonafide speaker labels from ASVspoof 2019 in hand. We unravel the efficacy of unsupervised domain adaptation methods with a PLDA trained with OOD data. As for the in-domain (InD) data, we explore with bonafide-only and bonafide-and-spoofed partitions (specified in Section~\ref{secsec:data}). Our back-end system with the methods is illustrated in Fig.~\ref{fig:methods}. The unsupervised DA methods covered in this paper are discussed below.

\textbf{Correlation Alignment} (CORAL). CORAL~\cite{coral} is a feature-based DA technique that aligns the covariance matrices of OOD features to compensate the InD ones by minimizing the distance between them. It aims at transforming the embedding space of the original data to the target one while preserving maximal similarity to the former. Each embedding is transformed as 
\begin{align}
\vec{\phi} \leftarrow \mathbf{C}_\text{I}^{\frac{1}{2}}\mathbf{C}_\text{o}^{-\frac{1}{2}}\vec{\phi},    
\end{align}
where $\mathbf{C}_\text{I}$ and $\mathbf{C}_\text{o}$ are the covariance matrices computed from InD and OOD data respectively. The transformation is done via \emph{zero-phase component analysis} (ZCA)~\cite{zca}. The transformed pseudo-InD embeddings are used to re-train the PLDA. CORAL has been applied to speaker verification in~\cite{coral_asv2018, coralplus, kaldi_aplda}.

\textbf{CORAL+}. Alternating from CORAL, CORAL+~\cite{coralplus} is a model-based method working on covariance matrices of the PLDA directly. Instead of replacing the covariance matrices $\mathbf{\Phi}_\text{b}$ and $\mathbf{\Phi}_\text{w}$ with ones computed from the transformed pseudo-InD data, CORAL+ updates them by linearly interpolating between the original and the pseudo-InD ones: 

\begin{align}
\begin{aligned}
    \mathbf{\Phi}_{b} &\leftarrow (1-\beta)\mathbf{\Phi}_{b} + \beta \mathbf{A}^{T}\mathbf{\Phi}_{b}\mathbf{A} \\
    \mathbf{\Phi}_{w} &\leftarrow (1-\lambda)\mathbf{\Phi}_{w} + \lambda \mathbf{A}^{T}\mathbf{\Phi}_{w}\mathbf{A}
\end{aligned}
\end{align}
with control parameters ${\beta, \lambda} \in [0,1]$. $\mathbf{A}=\mathbf{C}_\text{I}^{\frac{1}{2}}\mathbf{C}_\text{o}^{-\frac{1}{2}}$ is the transformation matrix. In order to enhance the uncertainty accounted and cover larger speaker subspace, CORAL+ performs \emph{simultaneous diagonalization} \cite{matrix_analysis} on both the original and pseudo-InD covariances. Details of CORAL+ can be found in~\cite{coralplus}. CORAL+ has been reported to outperform CORAL in DNN-based ASV~\cite{coralplus, kaldi_aplda}.

\textbf{Kaldi adaptation} (APLDA). The unsupervised PLDA adaptor from Kaldi toolkit \cite{kaldi_aplda} starts by performing eigen decomposition on the total covariance matrices $\mathbf{\Sigma}_{0}^{\frac{1}{2}}\mathbf{\Sigma}_{i}\mathbf{\Sigma}_{0}^{\frac{1}{2}} = \mathbf{P}\mathbf{\Delta} \mathbf{P}^{T} $, where $\mathbf{\Sigma}_{0} = \mathbf{\Phi}_\text{b,0} + \mathbf{\Phi}_\text{w,0}$ and $\mathbf{\Sigma}_{i} = \mathbf{\Phi}_\text{b,i} + \mathbf{\Phi}_\text{w,i}$ are covariance matrices computed from inD and inD and OOD data respectively.  The obtained eigenvalues (as diagonal matrix $\mathbf{\Delta}$) and vectors (as matrix $\mathbf{P}$) are then used to update the PLDA covariances, followed by simultaneous diagonalization. It shares operations in common with CORAL and CORAL+, except that it performs interpolation on diagonal values of the PLDA parameters via eigenvalues, instead of on the parameters themselves. Mathematical details and its application for DNN-based ASV can be found in~\cite{kaldi_aplda}. This is the method used in ASV system developed for ASVspoof 2019 challenge~\cite{asvspoof2019}.

\section{Dataset Description: ASVspoof 2019}
We conduct the experiments on ASVspoof 2019 corpus~\cite{wang2020asvspoof}. Originally launched in 2015, the biennial ASVspoof challenge\footnote{https://www.asvspoof.org/} series focus on assessing the vulnerability of ASV systems against different spoofing attacks and developing standalone countermeasures. The ASVspoof 2019 database furthers the achievements and protocol from its 2015 and 2017 predecessors, with a more controlled setup and evaluation protocols. It acquires more state-of-the-art neural-based algorithms on TTS and VC, as well as more careful simulation of replayed speech, which makes it also useful for fraud audio detection in real-time cases such as telebanking and smart homes. 

The database has two subsets: \emph{logical access} (LA) and \emph{physical access} (PA). The LA subset corresponds to a scenario where the attacks come in the form of synthetic and converted speech. Such speech cannot be perceptually detected by humans but can be distinguished by ASV systems if equipped with reliable models. State-of-the-art TTS algorithms were applied to construct the database, including but not limited to variational autoencoder (VAE) \cite{vae}, WaveCycleGAN \cite{wavecyclegan}, and Tacotron \cite{tacotron}. They were equipped by advanced vocoders such as WaveNet \cite{wavenet} and WORLD \cite{world_vocoder}, generating high-quality synthetic speech. The vocoders mentioned here were also used for VC. The PA subset corresponds to the case where the attacks are presented in a simulated physical space with varying positions of the speaker, microphone, and reverberation time. In the evaluation, two main factors are considered: distance between attacker and the speaker and the quality of the replaying device. 

Both the subsets have their dedicated training, development, and evaluation partition which are commonly used for assessing spoofing countermeasures. Apart from these, the dataset comes with separate enrollment files for ASV experiments. The audio data provided for training CM systems could be employed for ASV domain adaption. The summary of the protocols of the two subsets in terms of the number of trials is shown in Table \ref{tab:trial_stats}.

\begin{table}[htbp]
  \normalsize
  \centering
  \begin{tabular}{|c|cc|cc|}
    \hline
    \multirow{2}{*}{Trial Type} & \multicolumn{2}{|c|}{LA} & \multicolumn{2}{|c|}{PA} \\ 
    \cline{2-5}
     & dev & eval & dev & eval \\ \hline
    target & 1484 & 5370 & 2700 & 12960 \\ \hline
    bonafide non-target & 5768 & 33327 & 14040 & 123930 \\ \hline
    spoofed non-target & 22296 & 63882 & 24300 & 116640 \\ \hline
  \end{tabular}
\caption{Trial statistics for ASVspoof 2019.}
\label{tab:trial_stats}
\end{table}

\section{Experiments}
\label{sec:experiments}

\subsection{Data}
\label{secsec:data}
Our speaker embedding extractor is trained on the pooled training sets of VoxCeleb1 \cite{voxceleb1} and VoxCeleb2 \cite{voxceleb2} consisting of 7205 speakers. The OOD PLDA is trained on VoxCeleb1, with 1211 speakers. 

We use the various partitions of ASVspoof 2019 for domain adaptation. The bonafide training partition is the same bonafide data used for countermeasure training in \cite{asvspoof2019}. It contains 2580 and 5400 utterances for the LA and PA scenarios, respectively. The number of speakers is 20 in both scenarios. We use two subsets correspondingly: 1) \emph{bonafide}, which contains bonafide human speech only; 2) \emph{spoofed}, which is a collection of spoofed utterances corresponding to the same 20 speakers. The amount of data for the latter subset is $n$-times more than the former, where $n$ is the number of spoofing conditions.

\subsection{System Configuration}
In all experiments, we use 40-dimensional mel filterbanks with Hamming window as the acoustic features. The size and step size of the Hamming window is 25ms and 10ms, respectively, and the number of FFT bins is 512. We use \emph{x-vector} \cite{xvector2018} based on \emph{extended time-delayed neural network} (E-TDNN) \cite{etdnn}. Differently from \cite{etdnn}, we replace the statistics pooling layer with attentive statistics pooling \cite{astats_pooling} and employ \emph{additive angular softmax} \cite{aam_softmax} as the training loss. We extract the embedding for each input utterance from the first fully-connected layer after the pooling layer. 

The extracted vectors are centered, unit-length normalized, and projected with a 150-dimensional LDA, to train and adapt the PLDA. We adapt the OOD PLDA with the aforementioned DA methods, with the scaling factor of within-class and between-class covariances being set to be different for the two scenarios, following the protocol described in \cite{asvspoof2019}: $\alpha_\text{w}=0.25$, $\alpha_\text{b}=0.0$ for LA, and $\alpha_\text{w}=0.9$, $\alpha_\text{b}=0.0$ for PA. 

\subsection{Evaluation}

We create trials for both the LA and PA scenarios following \cite{asvspoof2019}. Additionally, we utilize two trials lists in each scenario: \emph{bonafide} trials includes a mix of bonafide target and zero-effort impostor trials", while \emph{spoofed} trials are composed by bonafide target and spoofed trial pairs, treated as impostors. Log-likelihood ratio (LLR) scores are produced for the trials and \emph{equal error rate} (EER) is used to gauge performance. As some of our analyses report EERs on per-attack bases, with a limited number of trials, we also report the parametric 95\% confidence intervals $(EER \pm \delta * Z_{\alpha/2}) * 100\%$, following methods described in \cite{confidence_interval_origin}. We set the related parameters $\delta=0.5\sqrt{\text{EER}(1-\text{EER})n_{+}+n_{-})/(n_{+}*n_{-})}$  and $Z_{\alpha/2} = 1.96$, where $n_{+}$ and $n_{-}$ are the number of target and nontarget trials, respectively.

\begin{figure*}[t]
  \begin{center}
    \centering
     {\includegraphics[width=17cm]{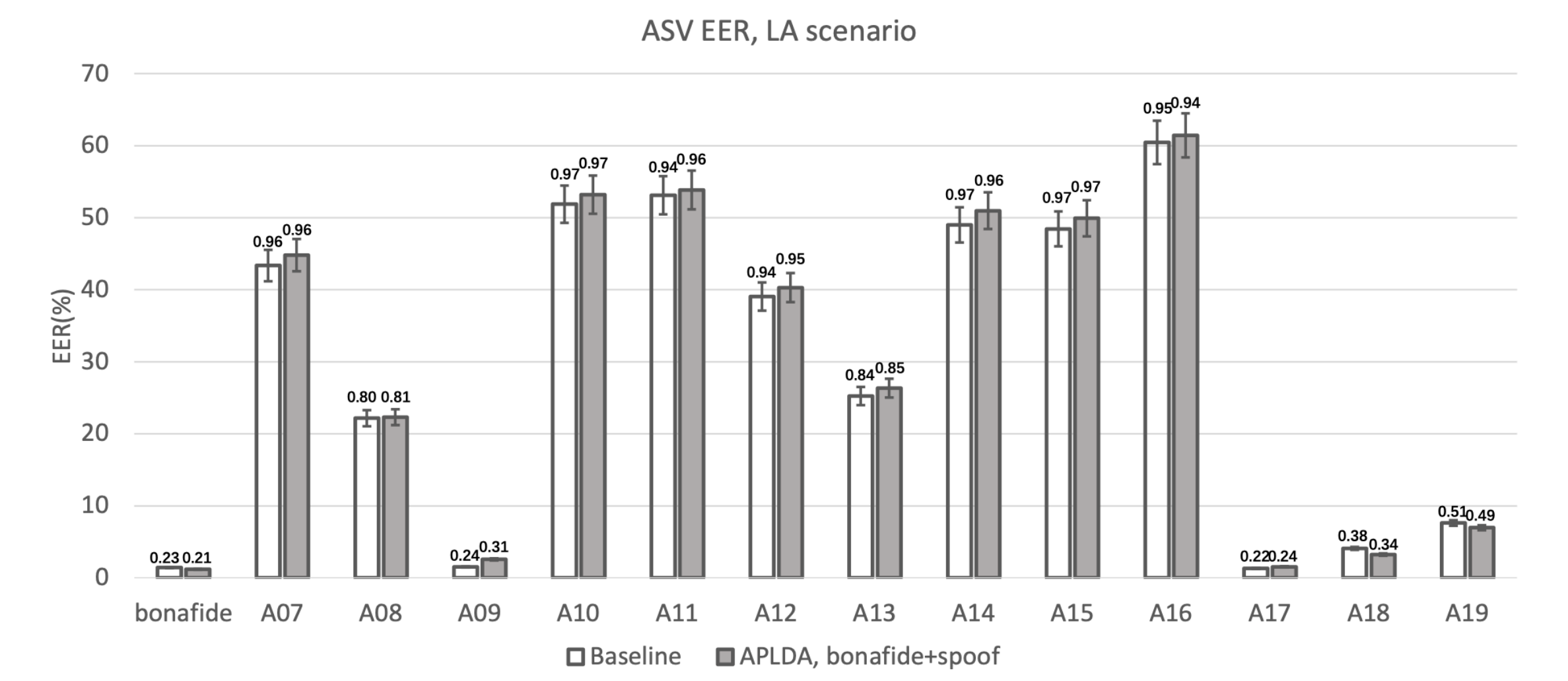}}
    \caption{Breakdown of ASV EER with baseline and best-performed method with respect to spoofing attacks for LA scenario. Numbers in small font size are the parametric confidence intervals.}
    \label{fig:asv_eer_la}
  \end{center}
\end{figure*}

\begin{figure*}[t]
  \begin{center}
    \centering
     {\includegraphics[width=18cm]{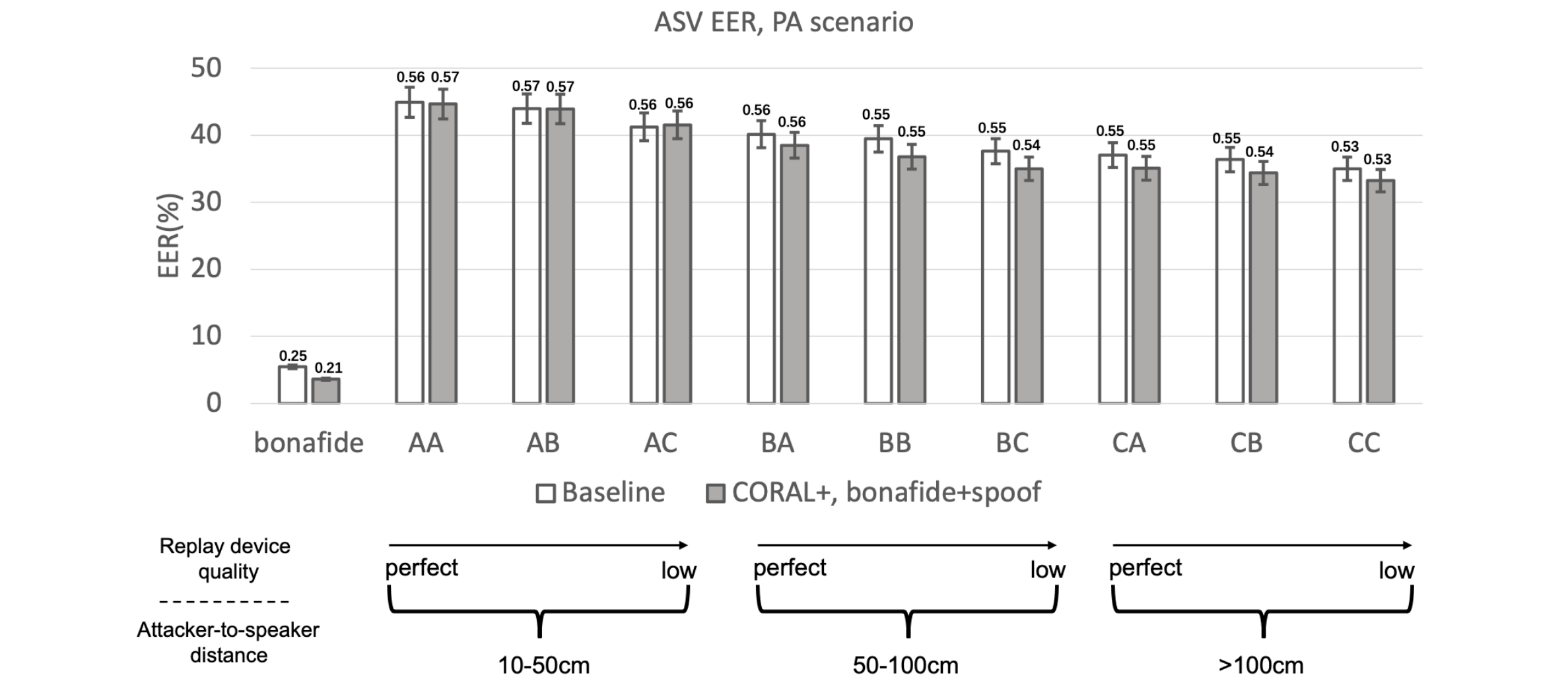}}, \caption{Breakdown of ASV EER with baseline and best-performed method with respect to spoofing attacks for PA scenario, along with the annotation of related variables \cite{asvspoof2019}. Numbers in small font size are the parametric confidence intervals.}
    \label{fig:asv_eer_pa}
  \end{center}
\end{figure*}

\begin{table}[htbp]
  \normalsize
  \centering
  \begin{tabular}{|c|c|cc|}
    \hline
    & & \multicolumn{2}{|c|}{ASV EER(\%)} \\ \hline
    Method & Data for adapt. & \emph{bonafide} & \emph{spoofed}
    \\ \hline
    -- & -- & 1.43 & \textbf{35.55} \\ \hline
    CORAL & \emph{bonafide} & \textbf{1.04} & 37.32 \\ \hline
    CORAL & \emph{bonafide+spoof} & 1.09 & 36.01 \\ \hline
    CORAL+ & \emph{bonafide} & 1.11 & 37.44 \\ \hline
    CORAL+ & \emph{bonafide+spoof} & 1.17 & 35.94 \\ \hline
    APLDA & \emph{bonafide} & 1.14 & 37.06 \\ \hline
    APLDA & \emph{bonafide+spoof} & 1.17 & 35.61 \\ \hline
  \end{tabular}
\caption{Results on ASVspoof 2019 LA evaluation.}
\label{tab:la_results}
\end{table}

\begin{table}[htbp]
  \normalsize
  \centering
  \begin{tabular}{|c|c|cc|}
    \hline
    & & \multicolumn{2}{|c|}{ASV EER(\%)} \\ \hline
    Method & Data for adapt. & \emph{bonafide} & \emph{spoofed} \\ \hline
    -- & -- & 5.46 & 39.95 \\ \hline
    CORAL & \emph{bonafide} & 3.75 & 39.48  \\ \hline
    CORAL & \emph{bonafide+spoof} & \textbf{3.49} & 39.07 \\ \hline
    CORAL+ & \emph{bonafide} & 3.86 & 39.41 \\ \hline
    CORAL+ & \emph{bonafide+spoof} & 3.61 & \textbf{37.85} \\ \hline
    APLDA & \emph{bonafide} & 4.11 & 39.79 \\ \hline
    APLDA & \emph{bonafide+spoof} & 3.79 & 39.27 \\ \hline
  \end{tabular}
\caption{Results on ASVspoof 2019 PA evaluation.}
\label{tab:pa_results}
\end{table}

\begin{table}[t]
  \normalsize
  \centering
  \begin{tabular}{|c|c|cc|}
    \hline
    & & \multicolumn{2}{|c|}{ASV EER(\%)} \\ \hline
    Method & Sampling & \emph{bonafide} & \emph{spoofed} \\ \hline
    APLDA & \emph{per-spk} & 1.12 & 37.17 \\ \hline
    APLDA & \emph{per-attack} & 1.17 & 37.13 \\ \hline
    APLDA& \emph{per-both} & 1.17 & 35.67 \\ \hline
  \end{tabular}
\caption{LA results of ablation study on data composition.}
\label{tab:la_ablation_study}
\end{table}

\begin{table}[ht]
  \normalsize
  \centering
  \begin{tabular}{|c|c|cc|}
    \hline
    & & \multicolumn{2}{|c|}{ASV EER(\%)} \\ \hline
    Method & Sampling & \emph{bonafide} & \emph{spoofed} \\ \hline
    CORAL+ & \emph{per-spk} & 4.07 & 39.31 \\ \hline
    CORAL+ & \emph{per-attack} & 4.11 & 39.25 \\ \hline
    CORAL+ & \emph{per-both} & 3.64 & 38.84 \\ \hline
  \end{tabular}
\caption{PA results of ablation study on data composition.}
\label{tab:pa_ablation_study}
\end{table}

\section{Results and Analysis}

\subsection{Logical Access}

Table \ref{tab:la_results} presents the results for the LA scenario. We first compare different methods with bonafide data for adaptation. In terms of bonafide EER, all the methods outperform the baseline (no adaptation), as expected. The maximum relative improvement of 27.3\% is provided by CORAL. Meanwhile, on EER from spoofed trials, the baseline achieves the lowest EER across all systems. APLDA with bonafide and spoof audios for adaptation reaches more comparable performance against the baseline.

We then focus on the effect of spoofed data for the adaptation. While including spoofed audios for adaptation increases bonafide EER across all methods, there are improvements in EER for spoofed trials. Maximum relative improvement on spoofed EER comes from CORAL+, by 4.1\%. Nevertheless, the increased amount of data does not outperform the baseline for the DA methods. Noting that LA corresponds to the cases where synthesized and converted speech segments are used for attacking, the results suggest that unsupervised DA methods may not be effective on generalization for synthesized spoofing attacks.

A breakdown of the results per attack is shown in Fig.~\ref{fig:asv_eer_la} for the best-performing system (APLDA). The baseline provides slightly lower EERs (though without significant differences) for most attacks. On attacks A18 and A19, APLDA outperforms baseline. Referring to \cite{asvspoof2019}, in these two attacks the spoofed audio is generated with VC via transfer learning from conventional statistical models such as GMM and i-vector, while other types of attack are mostly constructed via neural-based approaches. Similar observations can be found for other methods not shown in the figure as well. 

\subsection{Physical Access}

Table \ref{tab:pa_results} presents the results for the PA scenario. Focusing on systems with bonafide data for adaptation, notable improvements over the baseline are obtained. The maximum relative reduction on bonafide EER is 31.3\%, provided by CORAL. CORAL+ gives the lowest spoofed EER, lower than the baseline by relatively 1.4\%. 

Different from LA, augmenting adaptation data by spoofed utterances further improves the performance of all three DA methods. The lowest bonafide EER is achieved by CORAL,  with 36.1\% relative reduction over the baseline and 6.9\% relative reduction over the corresponding bonafide-only adaptation. Similar to LA, CORAL+ is efficient on spoofed evaluation trials, with 5.3\% relative EER reduction over the baseline. These observations suggest the potential of unsupervised adaptation on increasing awareness of the ASV system against computer-simulated room replay attacks. 

A break-down of the results per attack is shown in Fig. \ref{fig:asv_eer_pa} for CORAL+. Different from LA, there are more attacks where the DA method outperforms the baseline. Recall from \cite{asvspoof2019} that each attack is made of a duple, where the first letter refers to the distance between the attacker and the speaker (`A' is closest, `C' is farthermost) and the second letter refers to the quality of replay device (`A' denotes the best, `C' denotes the worst). The lowest EERs are obtained for attacks starting with `B' and `C', where the distance between the attacker and the talker is larger. In relative terms, these attacks are observed as being less detrimental. This indicates the potential of DA methods in handling attacks from more distant positions, while rather less affected by the quality of the replay device.

\subsection{Ablation Study on Data Composition}
Up to this point, we have compared different adaptation methods under two sets of adaptation data: bonafide only, and bonafide with spoofed data of bonafide speakers. 

A natural question on the amount of data arises: in the above settings, the amount of data between \emph{bonafide} and \emph{bonafide+spoof} are different. This is because the spoofing data are sampled separately and thus contain $n$ times more utterances than the bonafide data, where $n$ denotes the number of attacks present in the spoofed CM training set (6 for LA, 10 for PA). Thus, it is difficult to pinpoint whether performance differences are due to merely an increased amount of data, or rather, the type of data.

To address this question, we conduct an ablation experiment in both scenarios, where the total amount of bonafide and spoof data is held fixed and balanced (1290 utterances for both) but where we vary the composition of the spoof data. To compare with bonafide-only data setup, where assuming we have $m$ utterances, we sample $m/2$ utterances from the bonafide and spoofed set. For the bonafide part, we sample with the same size, with all speakers (20 of them) kept. We consider three alternative approaches: 
\begin{itemize}
    \item \emph{per-spk}: Sampling per speaker, where we sample $s_{1}=m/(2 \times 20)$ utterance for each speaker from the spoofed set;
    \item \emph{per-attack}: Sampling per attack, where we sample $s_{2}=m/(2\times n)$ utterances from the spoofed set for each attack;
    \item \emph{per-both}: The intersection of both the two factors. The total number of conditions here is $s_{1} \times s_{2}$ and we sample $n/(s_{1}\times s_{2})$ utterances from each condition. 
\end{itemize}
We separate the discussion concerning different scenarios and choose the best-performing system in terms of ASV EER on \emph{spoofed} trials (APLDA for LA, and CORAL+ for PA). 

The results are presented in Tables \ref{tab:la_ablation_study} and \ref{tab:pa_ablation_study}. For LA, compared to adaptation with only bonafide data, \emph{per-both} leads to the most competitive spoofed performance, with a relatively 18.1\% lower EER. This highlights the positive effect of spoofed data on enhancing the robustness of the system. Meanwhile, more spoofed data does not return significantly better performance on both trials, which indicates the relatively-low effect of the amount of spoofed data on improving the awareness of the system on synthetic speech.

For PA which corresponds to the replay attacks, replacing part of bonafide speech with spoofed ones does not lead to significantly better performance even on bonafide-only evaluation. The best one across different sampling schemes comes from the one covering all speakers and types of attack, outperforming adaptation with only bonafide data by relatively 1.4\%. The increasing amount of spoofed data this time leads to better performance, which is a different observation from LA.

\begin{table}[t]
  \normalsize
  \centering
  \begin{tabular}{|c|c|cc|}
    \hline
    & & \multicolumn{2}{|c|}{ASV+CM EER(\%)} \\ \hline
    Method & Data for adapt. & \emph{bonafide} & \emph{spoofed}
    \\ \hline
    -- & -- & 1.72 & 0.97 \\ \hline
    CORAL & \emph{bonafide} & 1.69 & 0.95  \\ \hline
    CORAL & \emph{bonafide+spoof} & 1.69 & 0.96 \\ \hline
    CORAL+ & \emph{bonafide} & 1.73 & 0.96 \\ \hline
    CORAL+ & \emph{bonafide+spoof} & 1.75 & 0.96 \\ \hline
    APLDA & \emph{bonafide} & 1.70 & 0.97 \\ \hline
    APLDA & \emph{bonafide+spoof} & 1.81 & 0.99 \\ \hline
  \end{tabular}
\caption{Results on LA, with integrated CM.}
\label{tab:cm_la_results}
\end{table}

\begin{table}[t]
  \normalsize
  \centering
  \begin{tabular}{|c|c|cc|}
    \hline
    & & \multicolumn{2}{|c|}{ASV+CM EER(\%)} \\ \hline
    Method & Data for adapt. & \emph{bonafide} & \emph{spoofed}
    \\ \hline
    -- & -- & 8.77 & 2.73 \\ \hline
    CORAL & \emph{bonafide} & 7.89 & 2.77  \\ \hline
    CORAL & \emph{bonafide+spoof} & 7.74 & 2.76 \\ \hline
    CORAL+ & \emph{bonafide} & 7.92 & 2.80 \\ \hline
    CORAL+ & \emph{bonafide+spoof} & 7.74 & 2.69 \\ \hline
    APLDA & \emph{bonafide} & 8.27 & 2.81 \\ \hline
    APLDA & \emph{bonafide+spoof} & 7.97 & 2.71 \\ \hline
  \end{tabular}
\caption{Results on PA, with integrated CM.}
\label{tab:cm_pa_results}
\end{table}

\subsection{Integration with Countermeasure System}
Up to this point, 
we have compared different adaptation methods in terms of their efficacy on 
a standalone ASV system. We would then like to investigate its integration with a fixed CM system.

We perform the parallel integration at score level using the Gaussian back-end fusion method described in \cite{parallel_cm_max2018}. 
It utilizes 2-dimensional vector $\mathbf{s} = [s_\text{CM}, s_\text{ASV}]^T$ consisting of CM score $s_\text{CM}$ and ASV score $s_\text{ASV}$. We define three classes: target, zero-effort impostor (aka non-target), and spoof impostor. Each of these three classes is modeled with a bivariate Gaussian. The 2D mean vectors and 2$\times$2 full covariance matrices of each class are obtained using their maximum likelihood estimates. Log-likelihood ratio scores for new trials are then computed by treating targets as the positive class (numerator) and the combined class of zero-effort impostors and spoof impostors as the negative class (denominator). Note that the latter is a two-component Gaussian mixture distribution. The mixing weight of each mixture component is set as $\alpha=0.5$. 
While the ASV scores correspond to one of the systems described above, the CM scores are produced using the method in \cite{lcnn}. We train \emph{light convolutional network} (LCNN) models for the LA and the PA scenarios using the respective training sets of ASVspoof 2019. 

The fusion results are displayed in Table \ref{tab:cm_la_results} and \ref{tab:cm_pa_results} for the LA and PA scenarios, respectively. As expected, fusion boosts the performance dramatically. On the LA scenario, the difference between the baseline and the DA methods reduces, regardless of the DA method or adaptation data. 
Similar observations can be found for PA, where the performance gap on the spoofed set is larger compared with LA, although slight degradation can be meanwhile observed for both scenarios. CORAL+ with both types of data for adaptation maintains the best performance on both bonafide and the spoofed evaluation sets across all systems. However, its relative improvement against the baseline is narrow. These findings indicate the gap between adaptation methods and separate CM ingredients, and the investigation of the former as open questions.

\subsection{Comparison with recent SASV challenge baseline}

Finally, we compare our results from the LA scenario with the (currently available) ASV results on the ongoing SASV challenge \cite{sasv_baseline_copied2022} (there is no PA scenario in this challenge) which shares the same data and protocols adopted for this study. Despite the shared evaluation setup, the compared ASV systems are very different. Thus, our aim is not a detailed discussion of the advantages or disadvantages of each architecture but, rather, reassurance that our results are reasonably well aligned with the (currently-reported) results of the SASV challenge.

This comparison is shown in Table \ref{tab:sasv_results}. Both our baseline and selected system (that reach the best performance on spoofed trial) return comparable performance against the SASV, with a relative improvement of 28.7\% in terms of bonafide EER provided by APLDA with bonafide and spoofed data for the adaptation. 

\begin{table}[t]
  \normalsize
  \centering
  \begin{tabular}{|c|cc|}
    \hline
    & \multicolumn{2}{|c|}{ASV EER(\%)} \\ \hline
    System & \emph{bonafide} & \emph{spoofed}
    \\ \hline
    ECAPA-TDNN \cite{sasv_baseline_copied2022} & 1.64 & 30.75 \\ \hline
    Baseline, no DA & 1.43 & 35.55 \\ \hline
    APLDA, \emph{bonafide+spoof} & 1.17 & 35.61 \\ \hline
  \end{tabular}
\caption{Comparison with SASV baseline.}
\label{tab:sasv_results}
\end{table}

\section{Conclusion}
We have conducted a preliminary study on spoofing-aware ASV with a special focus on unsupervised domain adaptation of a PLDA back-end. While our experiments also address fusion of ASV with a standalone spoofing countermeasure, our work is substantially differentiated from the majority of work on anti-spoofing that focuses on improving standalone countermeasures.

The key benefit of our unsupervised approach is simplicity. As the supply of spoofed speech data resources keeps increasing year by year, and since no architectural modifications to a conventional speaker recognition system is needed, it is straightforward to apply the technique to update existing PLDA back-end models for scenarios demanding increased spoof-awareness. 

While improvements on both bonafide and spoofed trials were obtained (especially in the PA scenario) through unsupervised domain adaptation, it is also evident that the absolute error rates on the spoofed trials remain too high on spoofing attacks. This may suggest that it is challenging to make a conventional speaker embedding extractor with PLDA back-end work on a mix of bonafide and spoofed data. Given the related activities in the ongoing SASV challenge \cite{sasv_baseline_copied2022}, we have have to reconsider either entirely different speaker embeddings, back-ends --- or both. 



\section{References}
{
\printbibliography
}

\end{document}